\documentclass[conference,a4paper]{IEEEtran}

\ifCLASSINFOpdf
\else
\fi
\usepackage[left=1.57cm,right=1.57cm,top=0.95cm,bottom=2.54cm]{geometry}
\usepackage{listings}
\usepackage{setspace}
\usepackage{graphicx}
\usepackage{tabularx,booktabs}
\usepackage{dingbat}
\usepackage{diagbox}
\usepackage{multirow} 
\usepackage[hyphens]{url}
\usepackage[hidelinks,breaklinks]{hyperref}
\usepackage{xcolor}
\usepackage{amsfonts, amsmath, amsthm, amssymb}
\usepackage{subcaption}
\hypersetup{breaklinks=true}
\usepackage{tikz}
\usepackage{textcomp}
\usepackage{lipsum}
\usepackage{svg}
\IEEEoverridecommandlockouts
\newcommand\copyrighttext{%
  \footnotesize \textcopyright 2026 IEEE.  Personal use of this material is permitted.  Permission from IEEE must be obtained for all other uses, in any current or future media, including reprinting/republishing this material for advertising or promotional purposes, creating new collective works, for resale or redistribution to servers or lists, or reuse of any copyrighted component of this work in other works.}
\newcommand\copyrightnotice{%
\begin{tikzpicture}[remember picture,overlay]
\node[anchor=south,yshift=10pt] at (current page.south) {\fbox{\parbox{\dimexpr\textwidth-\fboxsep-\fboxrule\relax}{\copyrighttext}}};
\end{tikzpicture}%
}
\usepackage{graphicx}
\graphicspath{{figures/}}
\begin{document}

%
\title{Evaluation of GPU Video Encoder for Low-Latency Real-Time 4K UHD Encoding}

\author{
    \IEEEauthorblockN{Kasidis Arunruangsirilert, Jiro Katto}
    \IEEEauthorblockA{Department of Computer Science and Communications Engineering, Waseda University, Tokyo, Japan
    \\\{kasidis, katto\}@katto.comm.waseda.ac.jp}
}
%

\maketitle

\copyrightnotice
\setstretch{0.92}
\begin{abstract}
The demand for high-quality, real-time video streaming has grown exponentially, with 4K Ultra High Definition (UHD) becoming the new standard for many applications such as live broadcasting, TV services, and interactive cloud gaming. This trend has driven the integration of dedicated hardware encoders into modern Graphics Processing Units (GPUs). Nowadays, these encoders support advanced codecs like HEVC and AV1 and feature specialized Low-Latency and Ultra Low-Latency tuning, targeting end-to-end latencies of \textless 2 seconds and \textless 500 ms, respectively. As the demand for such capabilities grows toward the 6G era, a clear understanding of their performance implications is essential. In this work, we evaluate the low-latency encoding modes on GPUs from NVIDIA, Intel, and AMD from both Rate-Distortion (RD) performance and latency perspectives. The results are then compared against both the normal-latency tuning of hardware encoders and leading software encoders. Results show hardware encoders achieve significantly lower E2E latency than software solutions with slightly better RD performance. While standard Low-Latency tuning yields a poor quality-latency trade-off, the Ultra Low-Latency mode reduces E2E latency to 83 ms (5 frames) without additional RD impact. Furthermore, hardware encoder latency is largely insensitive to quality presets, enabling high-quality, low-latency streams without compromise.

\end{abstract}



\setstretch{0.937}

%
\IEEEpeerreviewmaketitle

\vspace{-2.5mm}
\section{Introduction}

Low latency is critical for real-time applications, as high latency degrades the user's Quality of Experience (QoE) \cite{10.1145/3611070,8902462}. Achieving low latency is straightforward for low-throughput communications like VoIP and Online Games, but video's larger payload size presents a greater challenge. Transmitting uncompressed video is impractical for most users, as its bandwidth requirements exceed typical household internet capacity. Therefore, live video transmission necessitates sophisticated compression to be feasible over standard networks. These compression processes, however, introduce computational delay, creating a fundamental trade-off between transmission data rate, perceptual quality, and end-to-end latency.

To manage the large data volumes, lossy video compression algorithms compress videos to a manageable data rate for telecommunication infrastructures. These algorithms leverage spatial and temporal redundancies in the video stream and aspects of human visual perception, such as removing less critical high-frequency data through techniques like the Discrete Cosine Transform (DCT) and chroma subsampling. These processes, however, are computationally intensive. In optimized H.264/AVC systems, the encoder and decoder are the primary delay sources \cite{4588001}. While H.264/AVC provides adequate compression for High Definition (HD) video, the advent of Ultra High-Definition (UHD) presents new challenges by quadrupling the resolution of Full HD content \cite{9141900}. The growth in online video consumption and user-generated content, driven by platforms like TikTok and YouTube Shorts, has further strained Radio Access Network (RAN) infrastructures \cite{8992021}. Despite efforts to improve network capacity through new transmission techniques \cite{10570635,arunruangsirilert2025performanceanalysis5gfr2}, video traffic remains a major challenge, leading to issues such as legal disputes between ISPs and streaming services over explosive internet traffic \cite{lee_2021}. This necessitates the implementation of more efficient video codecs. \looseness=-1

Newer codecs such as High-Efficiency Video Codec (HEVC) \cite{6316136} and AV1 \cite{9363937} were introduced to address the growth of online video. However, their increased complexity demands greater computational resources, adding latency that can degrade QoE in applications like video conferencing and social Virtual Reality (VR) \cite{10.1145/3491101.3519678,10402741}. Software-based encoding of these codecs also consumes significant computational power and energy, making it impractical for battery-operated portable devices. In response, hardware solutions have been developed. Furthermore, the rise of online video game streaming prompted Graphics Processing Unit (GPU) manufacturers like Intel, AMD, and NVIDIA to integrate hardware video encoders onto their GPUs as early as 2011 \cite{7477508}. Modern GPU encoders now support recent codecs such as H.264/AVC, H.265/HEVC, and AV1. Our previous study found hardware encoders' RD performance approaches that of software encoders \cite{10637525}; however, it focused on maximum quality settings, evaluated only \textit{Normal Latency} tuning, and omitted the critical E2E latency metric.


The demand for high-resolution, low-latency video encoding has grown, particularly with the rise of video conferencing, cloud gaming and the needs of Smart Cities, Industrial Automation, and Autonomous Vehicles in the 6G era. While methods like predicting future video frames with deep learning have been explored to reduce latency \cite{10643859}, addressing the root cause of the delay is more effective. Consequently, hardware manufacturers have introduced \textit{Low-Latency} and \textit{Ultra Low-Latency} encoding modes on GPU encoders. Despite the importance of these modes for developing low-latency video communication systems, their effects are not yet well-understood. This paper evaluates the performance of these low-latency modes from both RD performance and End-to-End latency perspectives, comparing them to normal latency modes and software-based counterparts.

\vspace{-0.5mm}

\section{Experiment Setup}

\begin{table}[!tbp]
\setstretch{0.8}
\caption{Hardware and Software Configuration}
\vspace{-1.5mm}
\centering
\label{tab:hardware}
\resizebox{7.7cm}{!}{\begin{tabular}{@{}ll@{}}
\toprule
\multicolumn{2}{c}{NVIDIA and Intel GPU Encoding System}\\
\midrule
Hardware                 & Description  \\\midrule
System Board & ASUS ROG STRIX X570-F GAMING\\
CPU & AMD Ryzen 9 5900X 12-Core Processor \\
RAM & Dual-Channel DDR4 96 GB @ 3600 MHz \\ 
NVIDIA GPU & NVIDIA GeForce RTX 5070 Ti\\
Intel GPU & Intel Arc A770 Graphics\\\midrule
Software & Version \\\midrule
OS & Microsoft Windows 10 Pro Build 19045\\
ffmpeg & 2025-06-28-git-cfd1f81e7d-full\_build-www.gyan.dev\\
x264 & v0.165.3222 \\
x265 & 4.1-189-gcd4f0d6e9\\
SVT-AV1 & v3.0.2-83-g60c308f1 \\
NVIDIA GPU Driver & GeForce Game Ready Driver 572.42\\
Intel GPU Driver & 32.0.101.6913\\
Open Broadcaster Software & 31.1.1\\
\midrule
\multicolumn{2}{c}{AMD GPU Encoding System}\\
\midrule
Hardware                 & Description  \\\midrule
System Board & Dell Inspiron 14 7445 2-in-1\\
CPU & AMD Ryzen 7 8840HS \\
RAM & Dual-Channel DDR5 32 GB @ 5600 MHz \\ 
GPU & AMD Radeon 780M Graphics\\\midrule
Software & Version \\\midrule
OS & Microsoft Windows 11 Pro Build 26100\\
ffmpeg & 2025-06-28-git-cfd1f81e7d-full\_build-www.gyan.dev\\
AMD GPU Driver & 25.6.1\\
Open Broadcaster Software & 31.1.1\\\midrule
\multicolumn{2}{c}{WebRTC Server}\\
\midrule
Hardware                 & Description  \\\midrule
System Board & Dell PowerEdge R7515\\
CPU & AMD EPYC 7C13 64-Core Processor\\
RAM & Octal-Channel DDR4 512 GB @ 2666 MHz \\\midrule
Software & Version \\\midrule
OS & Ubuntu 24.04.1 LTS\\
WebRTC Server & SRS 6.0.166\\\midrule
\multicolumn{2}{c}{VMAF Calculation System}\\
\midrule
Hardware                 & Description  \\\midrule
CPU & AMD Ryzen Threadripper 3960X 24-Core Processor \\
RAM & Quad-Channel DDR4 128 GB @ 3200 MHz \\ 
GPU & 2x NVIDIA GeForce RTX 3090\\\midrule
Software & Version \\\midrule
OS & Ubuntu 22.04.5 LTS\\
ffmpeg & N-120252-g3ce348063c\\
libvmaf & v3.0.0 (b9ac69e6)\\
VMAF Model & vmaf\_4k\_v0.6.1neg \\
NVIDIA GPU Driver & 575.64.03 \\
NVIDIA CUDA Compiler & cuda\_12.9.r12.9\/compiler.36037853\_0 \\
\bottomrule
\end{tabular}}
\vspace{-7mm}
\end{table}
\vspace{-0.2mm}
\subsection{Rate-Distortion Performance Evaluation}


To evaluate the Rate-Distortion (RD) performance, two desktop GPUs from NVIDIA and Intel and one integrated GPU from AMD was chosen. Due to product availability, it was only possible to obtain the latest generation NVIDIA Blackwell Architecture GPU, while Intel and AMD GPU were the last generation products. While we acknowledge the use of different hardware generations, a key goal was to evaluate the most recent low-latency features available from each vendor at the time of testing. In our previous work \cite{10637525}, we found that unless new encoder feature is being added to the hardware encoder, the RD performance is usually nearly identical between generation, whether it is an integrated or a dedicated GPU. While this does not impact the results of Intel GPU, the hardware encoder on the latest AMD Navi IV Architecture GPU added B-frame encoding support for AV1 codecs, which may affect the RD performance.

The evaluation was conducted using 4K sequences from The Institute of Image Information and Television Engineers (ITE) Ultra-high definition/wide-color-gamut standard test sequences (Series A) \cite{ITE_2016}. This dataset comprises ten 4K sequences at 59.94 frames per second, conforming to the ITU-R standard. To align with typical live-streaming specifications, the 12-bit BT.2020 (HDR) source sequences were converted to 8-bit SDR with 4:2:0 chroma subsampling per the ITE method. To create a source compatible with GPU hardware decoders and accelerate the encoding process, this SDR dataset was then encoded into an All-Intra H.265/HEVC format using the \textit{x265} encoder with a medium preset and a Constant Rate Factor (CRF) of 10. All Video Multimethod Assessment Fusion (VMAF) and Peak Signal-to-Noise Ratio (PSNR) calculations were performed against this H.265/HEVC-encoded reference.

We determined appropriate bitrates by analyzing YouTube's VP9 transcodes of the ITE dataset. The analysis showed that YouTube's 4K VP9 transcodes ranged from 9.46 to 50.9 Mbps, averaging 29.5 Mbps. These figures were considered alongside the bitrates utilized for commercial UHD TV services in Japan and South Korea \cite{7522002}, which deliver 4K service via H.265/HEVC at approximately 25 Mbps and 29 Mbps, respectively. Based on this analysis, nine target bitrates were selected for the evaluation: 10, 15, 20, 25, 30, 35, 40, 45, and 50 Mbps. \looseness=-1

For this evaluation, three codecs were selected: H.264/AVC, H.265/HEVC, and AV1. The 'high' profile was used for H.264/AVC, while the 'main' profile was used for H.265/HEVC and AV1. The corresponding software encoders were x264, x265, and SVT-AV1, respectively. Each hardware and software encoder was tested across nine distinct configurations, derived from three latency-tuning modes and three encoding presets. As the Intel QuickSync API does not offer a latency tuning option, the parameters used for it follow the implementation of Open Broadcaster Software (OBS). Due to inconsistent naming conventions among manufacturers and software developers, the specific presets, tuning options, and corresponding FFmpeg parameters are detailed in Table \ref{tab:Parameter}. Since software encoding was conducted on the NVIDIA and Intel test systems, the software presets were chosen to ensure stable real-time encoding, ensuring a fair comparison against the primary use case for hardware encoders. Hence, the \textit{Quality} preset corresponds to the slowest setting that could be executed in real-time on the system CPU (see Table \ref{tab:hardware}).


\begin{table*}[!tbp]
\setstretch{0.77}
\caption{Encoding Preset and Latency Tuning for encoder each configuration with the corresponding FFmpeg parameters.}
\vspace{-1.5mm}
\centering
\label{tab:Parameter}
\resizebox{17.5cm}{!}{\begin{tabular}{@{}lccccc@{}}
\toprule 
\multicolumn{6}{c}{Encoding Preset}\\
\midrule
\multirow{2.5}{*}{Configuration} & \multirow{2.5}{*}{Manufacturer} & \multirow{2.5}{*}{Manufacturer's Name} & \multicolumn{3}{c}{FFmpeg Parameter}\\ 
\cmidrule(lr){4-6} &&& H.264/AVC & H.265/HEVC & AV1\\\midrule
\multirow{5.5}{*}{Quality}&AMD&Quality&-quality 2 -preset 2&-quality 0 -preset 0&-quality 30 -preset 30    \\\cmidrule(lr){2-6}
&Intel&TU1&-preset 1&-preset 1&-preset 1                                         \\\cmidrule(lr){2-6}
&NVIDIA&P7&-preset 18&-preset 18&-preset 18                                      \\\cmidrule(lr){2-6}
&Software&&-preset medium&-preset medium&-preset 8                               \\\midrule
\multirow{5.5}{*}{Balanced}&AMD&Balanced&-quality 0 -preset 0&-quality 5 -preset 5&-quality 70 -preset 70   \\\cmidrule(lr){2-6}
&Intel&TU4&-preset 4&-preset 4&-preset 4                                         \\\cmidrule(lr){2-6}
&NVIDIA&P4&-preset 15&-preset 15&-preset 15                                      \\\cmidrule(lr){2-6}
&Software&&-preset faster&-preset faster&-preset 10                              \\\midrule
\multirow{5.5}{*}{Speed}&AMD&Speed&-quality 1 -preset 1&-quality 10 -preset 10&-quality 100 -preset 100  \\\cmidrule(lr){2-6}
&Intel&TU7&-preset 7&-preset 7&-preset 7                                         \\\cmidrule(lr){2-6}
&NVIDIA&P1&-preset 12&-preset 12&-preset 12                                      \\\cmidrule(lr){2-6}
&Software&&-preset superfast&-preset superfast&-preset 12                        \\\midrule
\multicolumn{6}{c}{Latency Tuning}\\\midrule
\multirow{2.5}{*}{Configuration} & \multirow{2.5}{*}{Manufacturer} & \multirow{2.5}{*}{Manufacturer's Name} & \multicolumn{3}{c}{FFmpeg Parameter}\\ 
\cmidrule(lr){4-6} &&& H.264/AVC & H.265/HEVC & AV1\\\midrule
\multirow{5.5}{*}{Normal Latency}&AMD&transcoding&-usage 0 -pa\_lookahead\_buffer\_depth 0 -bf 3&-usage 0 -pa\_lookahead\_buffer\_depth 0&-usage 0 -pa\_lookahead\_buffer\_depth 0                                    \\\cmidrule(lr){2-6}
&Intel&Normal&-async\_depth 4 -look\_ahead\_depth 0 -bf 3&-async\_depth 4 -look\_ahead\_depth 0 -bf 3&-async\_depth 4 -look\_ahead\_depth 0 -bf 3                                    \\\cmidrule(lr){2-6}
&NVIDIA&High Quality&-tune hq -2pass 0  -multipass 0 -rc-lookahead 0&-tune hq -2pass 0  -multipass 0 -rc-lookahead 0&-tune hq -2pass 0  -multipass 0 -rc-lookahead 0        \\\cmidrule(lr){2-6}
&Software&&-rc-lookahead 0&-rc-lookahead 0&-svtav1-params "rc=2:pred-struct=1:lookahead=0"                                                                                  \\\midrule
\multirow{5.5}{*}{Low-Latency}&AMD&lowlatency&-usage 2 -pa\_lookahead\_buffer\_depth 0 -bf 0&-usage 2 -pa\_lookahead\_buffer\_depth 0 &-usage 1 -pa\_lookahead\_buffer\_depth 0                                         \\\cmidrule(lr){2-6}
&Intel&Low&-async\_depth 4 -look\_ahead\_depth 0 -bf 0&-async\_depth 4 -look\_ahead\_depth 0 -bf 0&-async\_depth 4 -look\_ahead\_depth 0 -bf 0                                       \\\cmidrule(lr){2-6}
&NVIDIA&Low Latency&-tune ll -2pass 0  -multipass 0 -rc-lookahead 0 -bf 0&-tune ll -2pass 0  -multipass 0 -rc-lookahead 0 -bf 0&-tune ll -2pass 0  -multipass 0 -rc-lookahead 0 -bf 0         \\\cmidrule(lr){2-6}
&Software&&-tune fastdecode -rc-lookahead 0 -bf 0&-tune fastdecode -rc-lookahead 0 -bf 0&-svtav1-params "rc=2:pred-struct=1:lookahead=0:fast-decode=1" -bf 0                \\\midrule
\multirow{5.5}{*}{Ultra Low-Latency}&AMD&ultralowlatency&-usage 1 -pa\_lookahead\_buffer\_depth 0 -bf 0 &-usage 1 -pa\_lookahead\_buffer\_depth 0 &-usage 2 -pa\_lookahead\_buffer\_depth 0                                    \\\cmidrule(lr){2-6}
&Intel&Ultra-Low&-async\_depth 1 -look\_ahead\_depth 0 -bf 0&-async\_depth 1 -look\_ahead\_depth 0 -bf 0&-async\_depth 1 -look\_ahead\_depth 0 -bf 0                                 \\\cmidrule(lr){2-6}
&NVIDIA&Ultra Low Latency&-tune ull -2pass 0  -multipass 0 -rc-lookahead 0 -bf 0&-tune ull -2pass 0  -multipass 0 -rc-lookahead 0 -bf 0&-tune ull -2pass 0  -multipass 0 -rc-lookahead 0 -bf 0\\\cmidrule(lr){2-6}
&Software&&-tune zerolatency -rc-lookahead 0 -bf 0&-tune zerolatency -rc-lookahead 0 -bf 0&-svtav1-params "rc=2:pred-struct=1:rtc=1:lookahead=0:fast-decode=2" -bf 0        \\
\bottomrule
\end{tabular}}
\vspace{-3mm}
\end{table*}

\begin{table*}[!tbp]
\setstretch{0.80}
\caption{Average Peak Signal-to-Noise Ratio (PSNR) and Video Multimethod Assessment Fusion (VMAF) scores for each encoder configuration, averaged across all test sequences and bitrates. Blue and red values indicate the best and worst performing encoder configuration for each codec and latency tuning combination. NL, LL, and ULL denote \textit{Normal}, \textit{Low}, and \textit{Ultra Low-Latency} Tuning. The "Diff." columns show the performance difference between the respective latency modes.}
\vspace{-1.5mm}
\centering
\label{tab:RDResults}
\resizebox{17.3cm}{!}{\begin{tabular}{@{}lc ccccc ccccc ccccc@{}}
\toprule 
\multicolumn{17}{c}{Average Peak Signal-to-Noise Ratio (PSNR) (dB)}\\
\midrule
\multirow{4}{*}{Manufacturer} & \multirow{4}{*}{Preset} & \multicolumn{15}{c}{Codec} \\ 
\cmidrule(lr){3-17} & &  \multicolumn{5}{c}{H.264/AVC} &\multicolumn{5}{c}{H.265/HEVC} & \multicolumn{5}{c}{AV1} \\
\cmidrule(lr){3-7}\cmidrule(lr){8-12}\cmidrule(lr){13-17} & & NL & LL & ULL & Diff. (NL-LL) & Diff. (LL-ULL)& NL & LL & ULL & Diff. (NL-LL) & Diff. (LL-ULL)& NL & LL & ULL & Diff. (NL-LL) & Diff. (LL-ULL)\\
\midrule
\multirow{3}{*}{AMD}&Quality&35.320&35.022&35.021&\textbf{-0.298}&-0.001&35.804&35.767&35.767&-0.037&0.000&36.138&36.138&36.133&0.000&-0.005    \\
&Balanced&35.270&34.972&34.967&\textbf{-0.298}&-0.005&35.797&35.759&35.759&-0.038&0.000&36.058&36.058&36.056&0.000&-0.002  \\
&Speed&35.231&34.934&34.928&\textbf{-0.297}&-0.006&\textcolor{red}{\textbf{35.437}}&\textcolor{red}{\textbf{35.396}}&35.396&-0.041&0.000&36.058&36.058&36.056&0.000&-0.002     \\\midrule
\multirow{3}{*}{Intel}&Quality&35.854&35.498&35.498&\textbf{-0.356}&0.000&36.388&\textcolor{blue}{\textbf{36.179}}&\textcolor{blue}{\textbf{36.179}}&\textbf{-0.209}&0.000&36.376&36.108&36.108&\textbf{-0.268}&0.000    \\
&Balanced&35.800&35.481&35.481&\textbf{-0.319}&0.000&36.344&36.096&36.096&\textbf{-0.248}&0.000&36.264&36.051&36.051&\textbf{-0.213}&0.000   \\
&Speed&35.782&35.469&35.469&\textbf{-0.313}&0.000&36.226&35.975&35.975&\textbf{-0.251}&0.000&36.207&35.991&35.991&\textbf{-0.216}&0.000      \\\midrule
\multirow{3}{*}{NVIDIA}&Quality&\textcolor{blue}{\textbf{35.984}}&\textcolor{blue}{\textbf{35.692}}&\textcolor{blue}{\textbf{35.692}}&\textbf{-0.292}&0.000&\textcolor{blue}{\textbf{36.391}}&36.120&36.120&\textbf{-0.271}&0.000&\textcolor{blue}{\textbf{36.588}}&\textcolor{blue}{\textbf{36.334}}&\textcolor{blue}{\textbf{36.334}}&\textbf{-0.254}&0.000    \\
&Balanced&35.907&35.669&35.669&\textbf{-0.238}&0.000&36.347&36.114&36.114&\textbf{-0.233}&0.000&36.569&36.284&36.284&\textbf{-0.285}&0.000   \\
&Speed&35.582&35.582&35.582&0.000&0.000&35.960&35.960&35.960&0.000&0.000&36.463&36.054&36.054&\textbf{-0.409}&0.000        \\\midrule
\multirow{3}{*}{Software}&Quality&35.004&\textcolor{red}{\textbf{34.009}}&34.796&\textbf{-0.995}&\textbf{0.787}&36.229&35.794&35.791&\textbf{-0.435}&-0.003&35.952&35.918&35.993&-0.034&0.075   \\
&Balanced&35.270&34.264&34.952&\textbf{-1.006}&\textbf{0.688}&36.134&35.859&35.741&\textbf{-0.275}&-0.118&\textcolor{red}{\textbf{35.474}}&\textcolor{red}{\textbf{35.474}}&\textcolor{red}{\textbf{35.680}}&0.000&\textbf{0.206}   \\
&Speed&\textcolor{red}{\textbf{34.909}}&34.245&\textcolor{red}{\textbf{34.526}}&\textbf{-0.664}&\textbf{0.281}&35.842&35.695&\textcolor{red}{\textbf{35.338}}&-0.147&\textbf{-0.357}&\textcolor{red}{\textbf{35.474}}&\textcolor{red}{\textbf{35.474}}&\textcolor{red}{\textbf{35.680}}&0.000&\textbf{0.206}      \\\midrule
\multicolumn{17}{c}{Average Video Multimethod Assessment Fusion (VMAF) No Enhancement Gain (NEG) 4K Score}\\
\midrule
\multirow{4}{*}{Manufacturer} & \multirow{4}{*}{Preset} & \multicolumn{15}{c}{Codec} \\ 
\cmidrule(lr){3-17} & &  \multicolumn{5}{c}{H.264/AVC} &\multicolumn{5}{c}{H.265/HEVC} & \multicolumn{5}{c}{AV1} \\
\cmidrule(lr){3-7}\cmidrule(lr){8-12}\cmidrule(lr){13-17} & & NL & LL & ULL & Diff. (NL-LL) & Diff. (LL-ULL)& NL & LL & ULL & Diff. (NL-LL) & Diff. (LL-ULL)& NL & LL & ULL & Diff. (NL-LL) & Diff. (LL-ULL)\\
\midrule
\multirow{3}{*}{AMD}&Quality&82.530&81.755&81.754&\textbf{-0.775}&-0.001&85.126&84.989&84.988&-0.137&-0.001&86.843&86.843&86.832&0.000&-0.011  \\
&Balanced&82.292&81.531&81.520&\textbf{-0.761}&-0.011&\textcolor{red}{\textbf{85.104}}&\textcolor{red}{\textbf{84.963}}&84.961&-0.141&-0.002&86.731&86.731&86.726&0.000&-0.005 \\
&Speed&82.054&81.352&81.339&\textbf{-0.702}&-0.013&85.631&85.463&85.462&-0.168&-0.001&86.731&86.731&86.726&0.000&-0.005    \\\midrule
\multirow{3}{*}{Intel}&Quality&85.021&83.863&83.863&\textbf{-1.158}&0.000&\textcolor{blue}{\textbf{86.878}}&\textcolor{blue}{\textbf{86.980}}&\textcolor{blue}{\textbf{86.980}}&0.102&0.000&86.502&86.554&86.554&0.052&0.000      \\
&Balanced&84.642&83.647&83.647&\textbf{-0.995}&0.000&86.803&86.792&86.792&-0.011&0.000&85.881&86.237&86.237&0.356&0.000    \\
&Speed&84.563&83.636&83.636&\textbf{-0.927}&0.000&86.492&86.558&86.558&0.066&0.000&85.641&86.053&86.053&0.412&0.000        \\\midrule
\multirow{3}{*}{NVIDIA}&Quality&\textcolor{blue}{\textbf{85.368}}&\textcolor{blue}{\textbf{84.930}}&\textcolor{blue}{\textbf{84.930}}&-0.438&0.000&86.354&86.473&86.473&0.119&0.000&\textcolor{blue}{\textbf{87.691}}&\textcolor{blue}{\textbf{87.463}}&\textcolor{blue}{\textbf{87.463}}&-0.228&0.000     \\
&Balanced&84.923&84.855&84.855&-0.068&0.000&86.159&86.434&86.434&0.275&0.000&87.636&87.395&87.395&-0.241&0.000    \\
&Speed&84.546&84.546&84.546&0.000&0.000&85.990&85.990&85.990&0.000&0.000&87.455&87.121&87.121&-0.334&0.000        \\\midrule
\multirow{3}{*}{Software}&Quality&82.340&79.670&81.395&\textbf{-2.670}&\textbf{1.725}&85.934&86.009&84.752&0.075&\textbf{-1.257}&85.819&85.799&85.989&-0.020&0.190    \\
&Balanced&82.174&\textcolor{red}{\textbf{79.019}}&81.282&\textbf{-3.155}&\textbf{2.263}&85.258&85.431&\textcolor{red}{\textbf{84.472}}&0.173&\textbf{-0.959}&\textcolor{red}{\textbf{85.424}}&\textcolor{red}{\textbf{85.424}}&\textcolor{red}{\textbf{85.644}}&0.000&0.220    \\
&Speed&\textcolor{red}{\textbf{80.796}}&79.466&\textcolor{red}{\textbf{79.758}}&\textbf{-1.330}&0.292&85.702&85.653&85.380&-0.049&-0.273&\textcolor{red}{\textbf{85.424}}&\textcolor{red}{\textbf{85.424}}&\textcolor{red}{\textbf{85.644}}&0.000&0.220      \\
\bottomrule
\end{tabular}}
\vspace{-6.5mm}
\end{table*}

Equivalent software parameters were determined by inspecting hardware encoder outputs with Elecard StreamEye 2024. Analysis of the hardware encoder output revealed \textit{Normal Latency} tuning used three B-frames. \textit{Low-Latency} tuning disabled B-frames, consistent with the software \textit{Fast Decode} tune, while \textit{Ultra Low-Latency}, which disabled parallelization and maintained a more consistent bit allocation, was analogous to the \textit{Zero Latency} tune. Therefore, for the evaluation, the \textit{Fast Decode} and \textit{Zero Latency} software tunings were selected as direct counterparts to the \textit{Low-Latency} and \textit{Ultra Low-Latency} hardware modes, respectively (see Table \ref{tab:Parameter}). \looseness=-1


Following recommendations from live-streaming platforms \cite{google, twitch}, all encodes utilized a two-second Group of Pictures (GOP) and Constant Bitrate (CBR) rate control, achieved by setting the target bitrate (\textit{-b:v}), max bitrate (\textit{-maxrate}), and buffer size (\textit{-bufsize}) to the same value for each test point (e.g., \textit{-b:v 25M -maxrate 25M -bufsize 25M}). The target, maximum, and minimum bitrates, as well as the buffer size, were set to the target value. Advanced features such as lookahead, two-pass encoding, Adaptive Quantization (AQ), and NVIDIA's Split Frame Encoding (SFE) \cite{nvidia_2024} were disabled. VMAF and PSNR scores were calculated against the H.265/HEVC reference using \textit{libvmaf\_cuda} with the \textit{vmaf\_4k\_v0.6.1neg} model to evaluate core codec performance and omit any enhancement gains.
\vspace{-0.5mm}
\subsection{End-to-End Latency Evaluation}

To evaluate the end-to-end latency, Open Broadcaster Software (OBS) was utilized. The H.265/HEVC software encoder (x265) could not be evaluated, as it was not implemented in OBS, and FFmpeg lacked support for the H.265/HEVC codec over the WebRTC-HTTP Ingestion Protocol (WHIP) at the time of this study. A source video displaying a 60 fps timecode was streamed at 25 Mbps from OBS to a local SRS server via WHIP, with encoding parameters configured for each test case. A separate client on the same network received the stream using the WebRTC-HTTP Egress Protocol (WHEP) via a web-based player. Latency was measured by capturing a side-by-side photograph of the source and playback monitors and calculating the difference in their displayed timecodes. To minimize extraneous delays, all systems were interconnected using Gigabit Ethernet on the same network switch, and monitors were connected directly to the output of the encoding and decoding GPU.
\vspace{-0.5mm}
\section{Results and Analysis}

\subsection{Rate-Distortion Performance}

The rate-distortion (RD) performance was evaluated using the average Peak Signal-to-Noise Ratio (PSNR) and Video Multimethod Assessment Fusion (VMAF) 4K scores, as presented in Table \ref{tab:RDResults} of the study.

For the H.264/AVC codec, transitioning from the \textit{Normal Latency} to the \textit{Low-Latency} mode resulted in an average PSNR decline of approximately 0.27 dB across all hardware encoders. This degradation is attributed to the disabling of B-frames in \textit{Low-Latency} mode, which reduces compression efficiency. An exception was the NVIDIA encoder at the Speed preset (P1), which produced identical outputs regardless of the latency tuning. In contrast, the x264 software encoder experienced a more substantial PSNR decline of 0.888 dB. While the PSNR drop was similar across hardware, VMAF degradation varied: NVIDIA's was smallest (0.168 points), followed by AMD (0.746 points), and Intel's was largest (1.027 points). The software encoder incurred a significant VMAF score loss of 2.385 points. Shifting from \textit{Low-Latency} to \textit{Ultra Low-Latency} mode did not introduce a significant drop in RD performance for hardware encoders. However, the \textit{zerolatency} tuning on the x264 encoder improved PSNR and VMAF by 0.585 dB and 1.426 points, respectively, compared to the \textit{fastdecode} tuning. For this codec, the NVIDIA encoder demonstrated the highest RD performance, followed by the Intel and AMD encoders, with the software encoder performing the worst.

In the case of the H.265/HEVC codec, enabling the \textit{Low-Latency} mode did not cause a significant RD performance drop for the AMD encoder. The Intel and NVIDIA encoders exhibited similar PSNR declines of 0.236 dB and 0.168 dB, respectively. Consistent with the H.264/AVC results, the NVIDIA encoder at the Speed preset (P1) produced identical output across all latency settings. The x265 software encoder also experienced a comparable PSNR drop of 0.286 dB. The VMAF scores indicated that switching to \textit{Low-Latency} mode did not substantially reduce RD performance, with minor gains and losses observed across encoders. Furthermore, transitioning to \textit{Ultra Low-Latency} tuning resulted in no change in either PSNR or VMAF for any hardware encoder. Conversely, the x265 encoder saw a notable decrease of 0.159 dB in PSNR and 0.830 points in VMAF. For this codec, the Intel encoder achieved the best RD performance, followed by NVIDIA and the x265 software encoder, while the AMD encoder performed the least efficiently. Further analysis revealed that the Intel encoder utilized Unidirectional B-frames instead of conventional P-frames, a behavior that could not be disabled despite setting \textit{-bf} flag to 0 and may affect compatibility with some playback devices.


Finally, for the AV1 codec, the AMD encoder produced identical output in both \textit{Normal Latency} and \textit{Low-Latency} modes, with the \textit{Ultra Low-Latency} mode yielding a nearly identical result. Similarly, the SVT-AV1 software encoder generated identical outputs for Presets 10 and 12. Unlike with other codecs, the NVIDIA AV1 encoder produced different outputs across all presets for each latency tuning, with an average PSNR and VMAF score drop of 0.316 dB and 0.268 points, respectively, when moving from \textit{Normal Latency} to \textit{Low-Latency}. The same transition on the Intel encoder led to a PSNR drop of 0.232 dB but a VMAF improvement of 0.273 points. On the other hand, the \textit{Ultra Low-Latency} configuration on the Software encoder resulted in a PSNR and VMAF score uplift of 0.163 dB and 0.210 points, respectively. For this codec, the NVIDIA encoder performed best, followed by the AMD and Intel encoders, while the software encoder had the lowest performance. \looseness=-1
\vspace{-0.8mm}
\subsection{End-to-End Latency}
\begin{table}[!tbp]
\setstretch{0.75}
\caption{Measured End-to-End latency in number of frames on 60p Timecode for each combination of resolution, codec, preset, and tuning. 'DNR' indicated that the configuration 'Did Not Run' or failed to achieve sufficient throughput for real-time encoding.}
\vspace{-1.5mm}
\centering
\label{tab:E2EResults}
\resizebox{8.4cm}{!}{\begin{tabular}{@{}lc ccc ccc ccc@{}}
\toprule 
\multicolumn{11}{c}{Measured End-to-End Latency (Number of Frames Delayed on 60p Timecode)}\\
\midrule
\multirow{4}{*}{Manufacturer} & \multirow{4}{*}{Preset} & \multicolumn{9}{c}{Codec} \\ 
\cmidrule(lr){3-11} & &  \multicolumn{3}{c}{H.264/AVC} &\multicolumn{3}{c}{H.265/HEVC} & \multicolumn{3}{c}{AV1} \\
\cmidrule(lr){3-5}\cmidrule(lr){6-8}\cmidrule(lr){9-11} & & NL & LL & ULL & NL & LL & ULL & NL & LL & ULL \\
\midrule
\multirow{3}{*}{AMD}&Quality&7&8&7&9&9&8&8&9&9        \\
&Balanced&7&8&8&8&8&8&6&7&7       \\
&Speed&7&7&9&7&7&8&7&7&6          \\\midrule
\multirow{3}{*}{Intel}&Quality&12&11&8&10&10&5&11&11&7  \\
&Balanced&11&11&8&10&10&5&11&10&6 \\
&Speed&11&10&8&10&10&5&11&10&5    \\\midrule
\multirow{3}{*}{NVIDIA}&Quality&37&7&7&\textcolor{red}{DNR}&7&7&7&7&7     \\
&Balanced&7&7&7&7&7&7&7&7&7       \\
&Speed&7&6&6&6&7&7&7&7&7          \\\midrule
\multirow{3}{*}{Software}&Quality&\textcolor{red}{DNR}&82&7&\multicolumn{3}{c}{Not Supported}&98&97&100    \\
&Balanced&61&59&7&\multicolumn{3}{c}{Not Supported}&85&89&88     \\
&Speed&61&41&7&\multicolumn{3}{c}{Not Supported}&90&87&90        \\
\bottomrule
\end{tabular}}
\vspace{-6.5mm}
\end{table}

E2E latency, measured in delayed 60p frames, is summarized in Table \ref{tab:E2EResults}. Hardware encoders consistently outperformed their software counterparts in latency as software encoders introduce delays ranging from 41 frames for \textit{Speed} encoding preset to over 90 frames for the higher-quality encoding preset with some failing to run in real-time completely. Hardware encoders generally kept E2E latency at or below 12 frames (200 ms), even with \textit{Normal Latency} tuning. This dropped to a minimum of 5 frames (83 ms) on the Intel encoder using the \textit{Ultra Low-Latency} mode for H.265/HEVC and AV1. Throughout testing, these hardware configurations demonstrated stable computational throughput, with no significant fluctuations in performance, utilization percentage or frame drops observed under the low-latency settings. \looseness=-1

Although the AMD encoder exhibited lower Rate-Distortion (RD) performance, it delivers fairly consistent E2E latency, achieving a delay of approximately 6-9 frames (100-150 ms) regardless of the preset or tuning mode. In contrast, the Intel encoder's \textit{Normal} and \textit{Low-Latency} tuning resulted in a slightly higher latency of 10-12 frames (167-200 ms). However, using its \textit{Ultra Low-Latency} tuning provided a significant improvement, reducing the delay to an average of 8, 5, and 6 frames for the H.264/AVC, H.265/HEVC, and AV1 codecs, respectively. As the Intel encoder delivers superior RD performance to AMD's, and its \textit{Ultra Low-Latency} tuning does not negatively impact RD performance compared to \textit{Low-Latency} tuning, it is a strong candidate for latency-sensitive applications.


On the other hand, NVIDIA encoder performed the most consistently, maintaining a latency of approximately 7 frames across nearly all presets and tuning modes. A notable exception occurred with the H.264/AVC codec under the \textit{Quality} preset (P7) and \textit{Normal Latency} tuning, where a non-disableable two-pass encoding enforcement increased latency to 37 frames. A second exception under the same settings with the H.265/HEVC codec caused encoder overload and dropped frames. This issue can be mitigated on high-end NVIDIA GPUs by enabling Split-Frame Encoding (SFE), which distributes the workload across multiple hardware encoders at the cost of a minor RD performance penalty. For the dual-encoder GPU used in this study, enabling SFE successfully resolved the overload issue for the P7 preset, restoring the E2E latency to 7 frames, similar to the rest.


The impact of encoding presets on latency was minimal for most hardware encoders. With the exception of some configuration on NVIDIA GPU, switching between \textit{Quality}, \textit{Balanced}, and \textit{Speed} presets typically altered the latency by only one frame, if at all. This suggests users can select higher-quality presets without a significant latency penalty. Similarly, the transition from \textit{Normal Latency} to \textit{Low-Latency} tuning provided only marginal improvements. The most substantial latency reduction was achieved with the \textit{Ultra Low-Latency} mode, which proved highly effective for minimizing delay, particularly for the Intel encoder.

\vspace{-1mm}
\section{Conclusions and Future Work}

In this paper, the \textit{Low-Latency} and \textit{Ultra Low-Latency} tuning of modern AMD, Intel, and NVIDIA GPU were evaluated in terms of RD performance and End-to-End (E2E) latency for real-time 4K UHD streaming, comparing them against standard hardware modes and software encoders. The results show that the hardware encoders significantly outperformed software counterparts in E2E latency, while performing slightly better in terms of Rate-Distortion (RD) performance. It was found that using \textit{Low-Latency} tuning does negatively impact RD performance of hardware encoders across the board, while yielding negligible E2E latency improvement. However, the \textit{Ultra Low-Latency} tuning proved most effective, with the Intel encoder achieving the lowest latency of 5 frames (83 ms) for H.265/HEVC and AV1 with no significant additional RD penalty compared to its standard \textit{Low-Latency} mode. Among vendors, NVIDIA's encoder offered the best balance of RD performance and consistent low latency. Intel provided the best H.265/HEVC RD performance and the lowest potential latency, though its use of non-standard structure may affect compatibility. AMD delivered predictable latency with comparatively lower RD performance. Finally, it was found that for hardware encoders, quality presets have a minimal impact on latency, allowing users to prioritize visual quality without a significant trade-off. We acknowledge that the use of different GPU generations may influence the results, and future work should aim to compare contemporaneous hardware. \looseness=-1


Future work will expand the evaluation to more diverse datasets (e.g., high-motion, HDR), assess performance under realistic network conditions (Beyond 5G and 6G), analyze power consumption for mobile/data center applications, and provide broader context by benchmarking against non-real-time presets and including subjective user studies (MOS).

\vspace{-0.5mm}
\section*{Acknowledgement}
\vspace{-0.5mm}
This paper is supported by the Ministry of Internal Affairs and Communications (MIC) Project for Efficient Frequency Utilization Toward Wireless IP Multicasting.




%
\setstretch{1}
\Urlmuskip=0mu plus 1mu\relax
\bibliographystyle{IEEEtran}
\bibliography{b_reference}

\end{document}